\documentclass[12pt]{article}
\usepackage{amsmath}
\usepackage{amssymb}
\usepackage{amsthm}
\usepackage{fullpage}
\usepackage[utf8]{inputenc}
\usepackage{mathtools}
\usepackage{url}
\usepackage[usenames,svgnames]{xcolor}

\usepackage[pagebackref=false]{hyperref} 
\usepackage[all]{hypcap} 

\theoremstyle{plain}

\newtheorem{theorem}{Theorem}[section]
\newtheorem{proposition}{Proposition}[section] 
\newtheorem{corollary}[theorem]{Corollary}

\newtheorem{remark}{Remark}[section]

\numberwithin{equation}{section} 

\hypersetup{
  breaklinks,colorlinks,
  citecolor=Green,
  linkcolor=BlueViolet,
  urlcolor=DarkBlue,
  pdftitle={Hypergeometric  $\tau$ - functions, Hurwitz numbers and enumeration of paths},
  pdfauthor={J. Harnad and A. Yu. Orlov},
}

\DeclareMathOperator{\tr}{tr}

\DeclareMathOperator{\stab}{stab}

\DeclareMathOperator{\cont}{cont}

\def\be{\begin{equation}}
\def\ee{\end{equation}}

\def\bea{\begin{eqnarray}}
\def\eea{\end{eqnarray}}

\def\bt{\begin{theorem}}
\def\et{\end{theorem}}

\def\bp{\begin{proposition}}
\def\ep{\end{proposition}}

\def\bc{\begin{corollary}}
\def\ec{\end{corollary}}

\def\br{\begin{remark}\rm\small}
\def\er{\end{remark}}

\def\ss{\subset}
\def\ra{\rightarrow}

\def\mt{{\mapsto}}

\def\&{&{\hskip -20pt}}

\def\FF{\mathcal{F}}
\def\JJ{\mathcal{J}}

\def\Cb{\mathbf{C}}

\def\Nb{\mathbf{N}}
\def\Pb{\mathbf{P}}

\def\Zb{\mathbf{Z}}

\begin{document}
\baselineskip 16pt
\medskip
\begin{center}
\begin{Large}\fontfamily{cmss}
\fontsize{17pt}{27pt}
\selectfont
\textbf{Hypergeometric $\tau$-functions, Hurwitz numbers and enumeration of paths}\footnote{Work of JH supported by the Natural Sciences and Engineering Research Council of Canada (NSERC) and the Fonds de recherche du Qu\'ebec - Nature et technologies (FRQNT);  work of AO supported by RFBR grant 14-01-00860.}
\end{Large}\\
\bigskip
\begin{large} {J. Harnad}$^{1,2}$ and {A. Yu. Orlov}$^{3, 4}$
 \end{large}
\\
\bigskip
\begin{small}
$^{1}${ \em Centre de recherches math\'ematiques,
Universit\'e de Montr\'eal\\ C.~P.~6128, succ. centre ville,
Montr\'eal,
QC, Canada H3C 3J7} \\
 e-mail: harnad@crm.umontreal.ca \\
\smallskip
$^{2}${ \em Department of Mathematics and
Statistics, Concordia University\\ 7141 Sherbrooke W.,
Montr\'eal, QC
Canada H4B 1R6} \\
\smallskip
$^{3}${ \em Institute of Oceanology, Nakhimovskii Prospekt 36, 
Moscow 117997, Russia} \\
\smallskip
$^{4}${\em National Research University, Higher School of Economics, \\
International laboratory of representation theory and mathematical physics \\
20 Myastnitskaya Ulitsa, 
Moscow 101001, Russia} \\
e-mail: orlovs@ocean.ru

\end{small}
\end{center}
\bigskip

\begin{abstract}
A multiparametric family of 2D Toda $\tau$-functions of hypergeometric type is
shown to provide  generating functions for composite,  signed Hurwitz numbers 
that enumerate certain classes of branched coverings of the Riemann sphere
and paths in the Cayley graph of $S_n$. The coefficients $F^{c_1, \dots, c_l}_{d_1, \dots,  d_m}(\mu, \nu)$
in their series expansion over products $P_\mu P'_\nu$ of power sum symmetric functions in
the two sets of Toda flow parameters and powers of the $l+m$ auxiliary parameters are shown
 to  enumerate $|\mu|=|\nu|=n$ fold branched covers of the Riemann sphere with specified ramification 
 profiles $ \mu$ and $\nu$ at a pair of points, and  two sets of additional branch paints, satisfying certain 
 additional conditions on their ramification profile lengths.  The first group  consists of  $l$ branch
 points,  with ramification profile  lengths  fixed to be the numbers   $(n-c_1, \dots, n- c_l)$; 
 the second consists of  $m$ further groups of  ``coloured''  branch points, of variable number,  
 for which the sums of the  complements of the ramification profile lengths within the groups are fixed to 
 equal the numbers $(d_1, \dots, d_m)$. The latter are counted with signs  determined by  the parity of the total number of
 such branch points. The coefficients $F^{c_1, \dots, c_l}_{d_1, \dots,  d_m}(\mu, \nu)$ 
 are also shown to enumerate paths in the Cayley graph of the symmetric group $S_n$  
 generated by transpositions, starting, as in the usual double Hurwitz case,  at an element  in the 
 conjugacy class of cycle type $\mu$ and ending in the class of type  $\nu$, with the first $l$ consecutive  
 subsequences of $(c_1, \dots, c_l)$ transpositions  strictly monotonically  increasing, and the 
 subsequent subsequences of $(d_1, \dots, d_m)$  transpositions weakly increasing. 
\end{abstract}

\section{Introduction}

  In \cite{GH} a general method for interpreting $2D$ Toda $\tau$-functions \cite{UTa, Ta, Takeb}  of hypergeometric type
  \cite{KMMM, OrSc} as combinatorial generating functions for certain classes of paths in the Cayley graph
  of the symmetric group $S_n$ was introduced. Examples included Okounkov's  generating function
  for double Hurwitz numbers \cite{Ok}, which count the number of inequivalent $n$-fold branched covers
  of the Riemann sphere $\Pb^1$ having a pair of branch points at $0$ and $\infty$
with  specified ramification profile types $\mu$ and $\nu$, and   $l$ additional branch points 
with simple  ramification type. The equivalent combinatorial interpretation is the enumeration 
of $k$-step paths in  the Cayley graph of $S_n$  generated by transpositions,  starting at an
 element in the conjugacy class of cycle type   $\mu$ and ending  in the class of cycle type $\nu$.   
  
    Several similar examples of   2D Toda $\tau$-functions of hypergeometric type
 were studied  in \cite{GGN1, GGN2, GH}  and interpreted   combinatorially in terms of counting   paths  
  in the  Cayley graph generated by transpositions  that are either  strictly  or weakly monotonically increasing, or some combination  thereof. These included several cases that, by restriction of  the
 flow variables to trace invariants of a pair of matrices, could be interpreted as matrix integrals of
 the Itzykson-Zuber-Harish-Chandra (HCIZ) type \cite{HC, IZ, GGN1, GGN2}, or variants thereof 
\cite[Appendix~A]{HOr1},  \cite{GH}.  In \cite{Z} a generating function  was given for Grothendieck's {\em dessins d'enfants}, 
 which is equivalent to the enumeration of  branched covers of Riemann surfaces with three branch points, or Belyi curves,
 one of which has specified ramification profile, and the other two specified profile lengths. This was subsequently 
 shown to  be a KP $\tau$-function, satisfying Virasoro constraints and topological recursion relations \cite{KZ, AMMN2}
and to have several equivalent representations as matrix integrals \cite{AC}. Other works concerned with
relating matrix models to coverings with three branch points include \cite{ Br,  MKR}. The relations between Hurwitz numbers
and Gromov-Witten invariants, together with the use of $\tau$-functions as generating functions 
for the latter  was further developed in \cite{OkP}.
 
 In \cite{OrSc} a large class  of  hypergeometric 2D Toda $\tau$-functions was studied, including a family that,
  when the flow variables are restricted to the trace invariants of a pair of $N\times N$ matrices, 
  can be interpreted as hypergeometric functions of matrix arguments  \cite{GR}. 
  (This was in fact  the origin of the term ``$\tau$-function of hypergeometric type''.) 
   It follows moreover, from the results of \cite{KMMM, Or, OrSh} and  \cite[Appendix~A]{HOr1}
   that these  may all be represented as matrix integrals.  In \cite{AMMN1, AMMN2} a  subclass of this
 family was noted to have  the form of generating  functions for Hurwitz numbers, but no general 
 combinatorial or geometric interpretation was given.  The combinatorial significance of the coefficients 
  in the double power sum symmetric function expansions for these was indicated  briefly in \cite{GH},  
  as counting  paths in the Cayley graph consisting of  $k$ strictly increasing subsequences of 
  transpositions having given lengths.
     
    In the present work, we view these as special cases of the more general class of hypergeometric 
    2 Toda $\tau$-functions introduced in \cite{OrSc}, and interpret them as generating
    functions  for the enumeration of certain classes of branched covers of $\Pb^1$, and certain
     paths in the Cayley graph of $S_n$ satisfying  specified geometric and combinatorial constraints.   
     The main results are stated in  Theorems \ref{geometric_interpretation}  and \ref{combinatorial_interpretation}, 
     which give both the  geometric  and combinatorial significance of the coefficients  
    $F^{(c_1, \dots , c_l)}_{(d_1, \dots, d_m)}(\mu, \nu)$  for the expansions of these $\tau$-functions
    in a basis of  products of power sum symmetric functions and  monomials in the additional parameters.
    
    Their  enumerative geometrical significance, given in  \autoref{geometric_interpretation}, 
  is that they count, with signs, the $n$-sheeted branched 
 covers of $\Pb^1$, having  again a pair of branch points  $(0, \infty)$,  with  ramification profiles
 given by a pair of partitions $(\mu,\nu)$, plus  two further families of branch points, 
  satisfying specific conditions.  The first family consists of $l$ branch points  whose ramification profiles 
 have specified lengths  $\{n- c_a\}_{a=1, \dots k}$.  The second consists of  $m$ ``coloured'' groups, each containing  
a variable number  of branch points, but constrained so that  the sums of the complements of the lengths of the
 ramification profiles  of the points  within each colour group  are equal to the specified numbers $\{d_b\}_{b=1, \dots m}$.
The sign in the counting is $(-1)^{mn +\sum_b^{m} d_b}$ times the parity of the total number of coloured branch points.
  
  The combinatorial meaning of the coefficients $F^{(c_1, \dots , c_l)}_{(d_1, \dots, d_m)}(\mu, \nu)$,
  given in  \autoref{combinatorial_interpretation}, is the following: these again enumerate 
 paths in the Cayley graph, starting  at an element in the  conjugacy class of cycle type $\mu$   
 and ending at one in the class of type $\nu$,  constrained so  that  the first $k$ consecutive subsequences 
 of transpositions of  lengths $(c_1 \dots , c_l)$ are each monotonically 
strictly increasing with respect to their larger elements,  while those in the  next successive subsequences
  of  lengths $(d_1, \dots , d_m)$  are weakly monotonically increasing. 
  
  All previously studied examples of generalized Hurwitz numbers can be recovered as special cases 
within this extended family of  combinatorial/geometric  generating functions.

\section[The 2D Toda $\tau$-functions $\tau^{(q, {\bf w}, {\bf z})}(N,{\bf t}, {\bf s})$]
{The hypergeometric 2D Toda $\tau$-functions $\tau^{(q, {\bf w}, {\bf z})}(N,{\bf t}, {\bf s})$}

\subsection[2D Toda $\tau$-functions of hypergeometric type]
{2D Toda $\tau$-functions of hypergeometric type }

A  2D Toda $\tau$-function \cite{UTa, Ta, Takeb} consists of a lattice of 
functions $\tau^{2D\text{T}}(N, {\bf t}, {\bf s})$,  labelled by the integers $N\in \Zb$,
depending differentiably on two infinite sequences of complex  flow variables
\be
{\bf t} =(t_1, t_2, \dots),  \quad {\bf s}= (s_1, s_2, \dots ),
\ee
and satisfying the infinite set of Hirota bilinear differential-difference equations, which
are constant coefficient bilinear differential equations in the $({\bf t}, {\bf s})$ variables, 
and finite difference equations in the lattice variable $N$. These can be concisely 
expressed through the following formal contour integral equality \cite{UTa, Ta, Takeb}
\bea
 &\&  \oint_{z=\infty}  z^{N'-N}e^{-\xi(\delta{\bf t}, z)} \tau^{2D\text{T}}(N,{ \bf t} +[z^{-1}] , {\bf s})
 \tau^{2D\text{T}}(N', { \bf t}+ \delta{\bf t}  - [z^{-1}],  {\bf s} + \delta{\bf s})  = \cr
 &\&  \oint_{z=0}  z^{N'-N} e^{-\xi(\delta{\bf s}, z^{-1})} \tau^{2D\text{T}}(N-1, {\bf t},   { \bf s} +[z])
 \tau^{2D\text{T}}(N'+1, {\bf t} + \delta{\bf t}, {\bf s}+ \delta{\bf s}  - [z] ) 
 \label{Hirota_bilinear}
 \eea
 for any pair $N, N' \in \Zb$, where
\be
\xi({\bf t}, z) :=\sum_{i=1}^\infty t_i z^i,  \quad   [z]_i  :=  {1 \over  i }z^i,
\ee
 understood as  satisfied  identically in the doubly infinite set of parameters
 \be
 \delta{\bf t}   = (\delta t_1, \delta t_2, \dots),\   \delta{\bf s} := (\delta s_1, \delta s_2, \dots) . 
 \ee

These imply, in particular, the full set of KP (Kadomtsev-Petviashvili) Hirota bilinear relations \cite{DJKM}
for either of the two sets of  flow variables ${\bf t}$ and ${\bf s}$, for each $N$, as well as
an infinite set of bilinear nearest neighbour difference equations linking the lattice sites $N, N'$ to their neighbours.
Such  general 2D Toda $\tau$-functions may be understood in terms of infinite abelian group
actions on infinite flag manifolds \cite{UTa, Ta}. Using the Pl\"ucker embedding, they
may be given a standard fermionic  Fock space representation 
\cite{Takeb} as  vacuum state expectation values. They may also be given an  infinite 
series representation as  sums over products $S_\lambda({\bf t}) S_\mu({\bf s})$ of Schur functions, in
which the coefficients are interpreted as Pl\"ucker coordinates of the associated infinite
flag manifold \cite{Takeb, HOr1}.
\br
To clarify notational conventions, when Schur functions are expressed in this way,
it is  understood that their arguments ${\bf t}$ and ${\bf s}$ are related  to
the monomial sum symmetric functions as follows:
\be
p_i = i t_i, \quad p'_i = i s_i,
\ee
and therefore  
\be
P_\mu({\bf t}) =  \prod_{i=1}^{\ell(\mu)}p_{\mu_i} =  \prod_{i=1}^{\ell(\mu)} \mu_i t_{\mu_i}. 
\ee
\er

 For the present work, only the special subfamily of $\tau$-functions of {\em hypergeometric type}
 \cite{KMMM, OrSc}  will be  needed, for which this expansion reduces to a diagonal sum over 
products $S_\lambda({\bf t}) S_\lambda({\bf s})$ of  Schur functions
of the same type. Further details and various applications of this subclass of 2D Toda $\tau$-functions
may be found in \cite{KMMM, OrSc, OrSh, HOr1, HOr2}. We give here only a brief summary of the essentials regarding 
the diagonal double Schur function  series representation as needed in the present work. 

For any map
\bea
\rho:\Zb &\& \ra \Cb^{\times} \cr
\rho: j &\&\  \mt \rho_j, 
\eea
we may define the following {\it content product} associated to the partition $\lambda$
\be
r_\lambda(N) := r_0(N) \prod_{(ij) \in \lambda} r_{N+j-i}, 
\label{r_content_product}
\ee
where
\be
r_j:= {\rho_j \over \rho_{j-1}}, \quad  r_0(N) := \prod_{j=0}^{N-1}\rho_j ,\quad  r_0(0) := 1,
\quad r_0(-N) = \prod_{i=1}^N \rho_{-j}^{-1},   \quad N \in \Nb^+.
\ee
As will be detailed further in Sections 4 and 5,  this may be viewed as the eigenvalues of a certain family of
operators acting either on the  direct sum  $\oplus_{n=1}^\infty \Zb(\Cb[S_n])$ of the centers of the group algebras 
of $S_n$, $n\in \Nb$ or, equivalently,  on  a fermionic  Fock space $\FF$, and used thereby
to define a 2D Toda  $\tau$-function of  hypergeometric type.  This may be expressed as a formal 
diagonal sum over products of Schur functions
\be
\tau_r(N, {\bf t }, {\bf s}) := \sum_{\lambda} r_\lambda(N) S_\lambda({\bf t}) S_\lambda({\bf s}), 
\label{hypergeometric_tau_expansion}
\ee
where $S_\lambda$ denotes the Schur function labelled by  the integer partition $\lambda= (\lambda_1 \ge \cdots \lambda_{\ell(\lambda)} > 0, 0 \dots)$ of length $\ell(\lambda)$, and weight $|\lambda|= \sum_{i=1}^{\ell(\lambda)}\lambda_i$.
It follows from the fermionic representation (see \cite{Takeb, OrSc} and  \autoref{fermionic_representation})
that any such lattice of functions $\tau_r(N, {\bf t }, {\bf s})$ satisfies the Hirota bilinear relations (\ref{Hirota_bilinear}). 

In the following, we assume some familiarity with  properties of the algebra $\Lambda$ 
of symmetric functions in an arbitrary number of variables \cite{Mac},  the group algebra $\Cb[S_n]$, and 
 irreducible characters $\chi_\lambda(\mu)$ of $S_n$   \cite{FH}.
The Frobenius character formula expresses the Schur functions $S_\lambda$  linearly in terms of the
power sum symmetric functions
  $\{P_\mu\}$, 
  \be
S_\lambda= \sum_{\mu \atop |\mu|=|\lambda|} {\chi_\lambda(\mu) P_\mu \over Z_\mu}, 
\label{frobenius_character}
\ee
where  $\chi_\lambda(\mu)$  is the irreducible character of $S_n$ corresponding to the 
partition $\lambda$, evaluated on the conjugacy class of cycle type given by the partition $\mu$, and
\be
Z_\mu := \prod_{i=1}^n i^{j_i}( j_i)!  =  |\stab(\mu)|,  \quad( j_i = \text{number of parts  of } \mu \text{ equal to } i),
\ee 
 is the order  of the stabilizer of any of the elements of the conjugacy class.
  
 Substituting this into  formula (\ref{hypergeometric_tau_expansion}) gives an expansion over products of 
 pairs of monomial sum symmetric functions
 \be
 \tau_r(N, {\bf t}, {\bf s})= \sum_{\mu, \nu \atop |\mu|= |\nu|} G_r(\mu, \nu) P_\mu({\bf t}) P_\nu({\bf s})
 \ee
 where
 \be
 G_r(\mu, \nu) =
 ( Z_\mu Z_\nu)^{-1}  \sum_{\lambda \atop |\lambda|=|\mu|=|\nu|} r_\lambda(N) \chi_\lambda(\mu) \chi_\lambda(\nu). 
 \label{G_mu_nu}
 \ee
 
 We also use the notation
\be
h_\lambda = {|\lambda|! \over d_\lambda}
\ee 
to denote  the product of hook lengths  \cite{Mac}, where 
\be 
d_\lambda = \chi_\lambda (1^{|\lambda|} ) = |\lambda|!  \det\left ({1 \over (\lambda_i -i +j)! }\right)
\ee
is the dimension of the irreducible representation of $S_n$ with character $\chi_\lambda(\mu)$.

\subsection[The family of  hypergeometric $\tau$-functions $\tau^{(q, {\bf w}, {\bf z}) }(N, {\bf t}, {\bf s})$]
{The family of hypergeometric $\tau$-functions  $\tau^{(q, {\bf w}, {\bf z} )}(N, {\bf t}, {\bf s})$ }

Choosing a set of $1+  l+m $ complex parameters $(q, {\bf w}, {\bf z})$,
\be
 \quad {\bf w} := (w_1, \dots, w_l), \quad {\bf z} := (z_1, \dots , z_m), 
\ee 
 Let $ \rho^{(q, {\bf w}, {\bf z})}_0 = 1$ and, for $j>0$,
\bea
\rho^{(q, {\bf w}, {\bf z})}_j &\&:=q^j \prod_{a=1}^l\prod_{b=1}^m \prod_{k=1}^j\left( 1+ k w_a \over 1-k z_b\right),  \\
 \rho^{(q, {\bf w}, {\bf z})}_{-j}&\&:= q^{-j} \prod_{a=1}^l\prod_{b=1}^m
 \prod_{k=0}^{j-1}\left( 1+ k z_b\over 1-k w_a\right).
  \label{rho_q_w_z_def} 
 \eea
 Then 
 \be
r_j ^{(q, {\bf w}, {\bf z})}  := q  \prod_{a=1}^l \prod_{b=1}^m  \ \left( { 1+  j w_a \over  1-  j z_b }\right)
\quad  \text{for all  } j\in \Zb
\ee
and
\be
r_\lambda^{(q, {\bf w}, {\bf z})}(N)  = r_0^{(q, {\bf w}, {\bf z})} (N) \prod_{(i,j)\in \lambda} r_{N+j-i}^{(q, {\bf w}, {\bf z})} , 
\label{content_product_q_c_d}
\ee
where
\bea
r_0^{(q, {\bf w}, {\bf z})}(N) &\&= q^{\frac{1}{2}N(N-1)} \prod_{j=1}^{N-1}  \left( {1 + (N-j)w_a \over 1 - (N-j)z_b} \right)^j, \quad N >  0,
\quad r_0(0) =1, \\
r_0^{(q, {\bf w}, {\bf z})}(-N) &\& = q^{\frac{1}{2}N(N+1)} \prod_{j=1}^{N}  \left( {1 - (N-j)w_a \over 1 +  (N-j)z_b} \right)^j,  \quad N >  0.
\eea

The resulting hypergeometric 2D Toda $\tau$-function is denoted
\bea
\tau^{(q, {\bf w}, {\bf z})} (N, {\bf t}, {\bf s}) &\&=
\sum_{\lambda} r_\lambda^{(q, {\bf w}, {\bf z})}(N)  S_\lambda({\bf t}) S_\lambda({\bf s})
\label{hurwitz_hypergeometric_tau_S}\\
&\&=  \sum_{\lambda} r_\lambda^{(q, {\bf w}, {\bf z})}(N)
 \sum_{\mu, \nu, |\mu| =|\nu|=n}   (Z_\mu Z_\nu)^{-1}\chi_\lambda(\mu) \chi_\lambda(\nu) P_\mu({\bf t} )P_\nu ({\bf s})  . \cr
&\&
\label{hurwitz_hypergeometric_tau_P}
\eea

\br
Since  $\tau^{(q, {\bf w}, {\bf z})} (N, {\bf t}, {\bf s})$  can be expressed in terms of
$\tau^{(q, {\bf w}, {\bf z})} (0, {\bf t}, {\bf s}) $  by a simple transformation of parameters
\be
\tau^{(q, {\bf w}, {\bf z})} (N, {\bf t}, {\bf s})  = T(q, N,  {\bf w}, {\bf z}) \tau^{(\tilde{q}, \tilde{\bf w}, \tilde{\bf z})} (0, {\bf t}, {\bf s}), 
\ee
where
\be
\tilde{w}_a:={w_a \over (1 + N w_a )}, \quad  \tilde{z}_b := {\tilde{z}_b  \over (1 - N z_b )}, 
\quad  \tilde{q} = q {\prod_{a=1}^l (1+N w_a )\over \prod_{b=1}^m (1- Nz_b)} , 
\ee
and
\be
T(q, N,  {\bf w}, {\bf z}) =  q^{{1\over2}N(N-1)}\prod_{a=1}^l \prod_{b=1}^m 
\left(  {1+Nw_a \over 1- Nz_b }\right)^{{1\over 2} N(N-1)}
\ee
we henceforth only consider the case $N=0$,  and simplify the notation to
\be
\tau^{(q, {\bf w}, {\bf z})} (0, {\bf t}, {\bf s}) =: \tau^{(q, {\bf w}, {\bf z})} ( {\bf t}, {\bf s}),
\quad  r_\lambda^{(q, {\bf w}, {\bf z})}(0) =: r_\lambda^{(q, {\bf w}, {\bf z})}.
\ee

\er

\br
If a positive integer $M>0$  is chosen, and the flow variables $({\bf t}, {\bf s})$
are restricted to be the trace invariants
\be
t_i ={1\over i}\tr (X^i) :=[X]_i,  \quad s_i = {1\over i}\tr (Y^i) := [Y]_i
\ee
of a pair $(X,Y)$ of $M\times M$ hermitian matrices whose eigenvalues are 
$\{x_i\}_{i=1,\dots , M}$, $\{y_j\}_{j=1, \dots, M}$, respectively, 
the series (\ref{hurwitz_hypergeometric_tau_S}) for $N=0$ only involves sums  over Schur functions
$S_\lambda$ for which $\ell(\lambda)\le M$, and may be related to the so-called
hypergeometric function of matrix arguments \cite{GR, OrSc} as follows.
We may consistently choose $z_1=-{1 \over M}$, since the restriction 
 $\ell(\lambda) \le M$ implies that there is no value of $j$ appearing in which
the denominator factor  in $r_j^{(q, {\bf w}, {\bf z})}$ vanishes.
Choosing the remaining  parameters to be nonvanishing,  and defining
\be
u_a=\frac {1}{w_a},\, a=1,\dots, l \ \text{  and }\  v_{b-1}:=-\frac{1}{z_b},\, b=2,\dots,m, 
\ee 
we have
\bea
\tau^{(q, {\bf w}, {\bf z})}(0, [X], [Y])
=&\&{ _{l}\Phi}_{m-1}(u_1,\dots, u_l; v_1,\dots, v_{m-1}|qX,Y)  \cr
&\& \cr
:= &\&{ \det \left({_{l}F}_{m-1}(u_1,\dots, u_l; v_1,\dots, v_{m-1}|q x_iy_j)\right)_{i, j=1,\dots,m} \over \Delta({\bf x}) \Delta({\bf y})}, 
 \eea
 where ${_{l}F}_{m}$ is the usual general hypergeometric function
\be
 {_{l}F}_{m}(u_1,\dots, u_l; v_1,\dots, v_{m}|x)=\sum_{n=0}^\infty \frac{\prod_{a=1}^l \,(u_a)_n}{\prod_{b=1}^{m} \,(v_b)_n} \,\frac{x^n}{n!}. 
\ee
Here ${_{l}\Phi}_{m}(a_1,\dots, a_l; v_1,\dots, v_{m}|X,Y) $ is what is known as the hypergeometric function 
of two matrix arguments \cite{GR}.

\er

Further expanding the coefficients in the formula (\ref{hurwitz_hypergeometric_tau_P}) for  
$\tau^{(q, {\bf w}, {\bf z})} ( {\bf t}, {\bf s})$ as  power series in the parameters $(q, {\bf w}, {\bf z})$, using multi-indices ${\bf c} = (c_1, \dots, c_l) \in \Nb^l$, ${\bf d} = (d_1, \dots, d_m) \in \Nb^m$, gives
\be
\tau^{(q, {\bf w}, {\bf z})} ( {\bf t}, {\bf s}) =
\sum_{n=0}^\infty q^n  \sum_{\mu, \nu \atop |\mu|=|\nu|=n}  \sum_{{\bf c}\in \Nb^l} 
\sum_{{\bf d}\in \Nb^m}  {\bf w}^{\bf c}   {\bf z}^{\bf d}   F^{\bf c}_{\bf d}(\mu, \nu)
P_\mu({\bf t})P_\nu({\bf s}).
\label{double_power_sum_exp}
\ee

We are now ready to state the two main theorems:

\noindent
\bt{\bf Geometric interpretation: generalized Hurwitz numbers.} 
\label{geometric_interpretation}
The coefficients  $ F^{\bf c}_{\bf d}(\mu, \nu)$ in the expansion (\ref{double_power_sum_exp})
are equal to  the  number of  $n$-sheeted inequivalent branched coverings  of the Riemann  
sphere  by a surface of  genus $g$ given by the Riemann-Hurwitz formula
\be
2g =2+  \sum_{a=1}^l  c_a+\sum_{b=1}^m d_b - \ell(\mu) - \ell(\nu),
\label{riemann_hurwitz}
\ee
counted with signs, as indicated below, such that the branch
points consist of three classes:

\noindent i)  A pair of branch points $(0,  \infty)$, with ramification profiles $(\mu, \nu)$.

\noindent ii) A set of $l$ further branch points $\{q_a\}_{a=1, \dots, l}$ ,
with ramification profiles $\{ \mu^{(a)}\}_{a=1, \dots, l}$, the complement of whose lengths  are
\be
n- \ell(\mu^{(a)}) = c_a,  \quad a=1, \dots, l
\ee

\noindent iii) A set of  $m$ further groups of branch points, $\{p_{b, i_b}\}_{b=1, \dots m \atop i_b =1, \dots j_b}$ ,  
 labeled by ``colours'' $b=1, \dots m$,  with  ramification profiles $\{ \nu^{(b,i_b)}\}_{b=1, \dots, m; i_b=1, \dots j_b}$ , 
 where $j_b$ is the number  of points in the $b$th coloured
  group,  such that the sum of the complements of the  lengths of the ramification  profiles
   at the points  $\{p_{b, i_b}\}_{i_b=1, \dots j_b}$ within the $b$th group is equal to $d_b$
    \be
  \sum_{i_b=1}^{j_b}\left(n -\ell(\nu^{(b,i_b}) \right)= d_b, \quad b= 1, \dots ,m.
  \label{summed_length_complement}
  \ee
 Each such covering is counted with a sign  $(-1)^{mn + C + D} $, where
 \be
C := \sum_{b=1}^m j_b,
\ee
is the total number of coloured branched points and
\be
 \quad D:=\sum_{b=1}^m  d_b  = nC -   \sum_{b=1}^m\sum_{i_b=1}^{j_b}\ell(\nu^{(b, i_b)})
\ee 
is the sum of the complements of the lengths of the ramification profiless of the coloured branch points. 

\et

\br
Note that the number of branch points in each of the  groups is variable,
but the sum is only over those in group (ii)  for which the lengths  $\ell(\mu^{(a)})$
are fixed and those in group (iii) for which the sum of the complements of the  lengths $D$ are fixed.
These therefore only involve signed sums over a finite number of individual Hurwitz numbers. It is also understood
that,  when applied to coverings by orientable surfaces,  this interpretation of $F^{\bf c}_{\bf d}(\mu, \nu)$,
is  valid only  if the genus $g$ given by formula (\ref{riemann_hurwitz}) is an integer. The case when it is a
half integer is applicable to counting nonorientable covers.
\er

\bt{\bf Combinatorial interpretation: multimonotonic paths in the Cayley graph}.
\label{combinatorial_interpretation}
The coefficients  $ F^{\bf c}_{\bf d}(\mu, \nu)$ in the expansion (\ref{double_power_sum_exp})
are equal to the number of paths in the Cayley graph of $S_n$ generated by transpositions $(a\, b)$, 
$a<b$,  starting at an element in the conjugacy class with cycle type given by the partition $\mu$
and ending in the conjugacy class with cycle type given by partition  $\nu$, such that the paths
consist of a sequence of 
\be 
 k:= \sum_{a=1}^l c_a + \sum_{b=1}^m d_b
 \ee
 transpositions $(a_1b_1) \cdots (a_k b_k)$, divided into $l+m$ subsequences, the first $l$ of which
consist of  $\{c_1, \dots, c_l\}$ transpositions that are strictly monotonically increasing
(i.e.\ $ b_i < b_{i+1}$ for each neighbouring pair of transpositions within the subsequence),
 followed by  $\{d_1, \dots, d_m\}$ subsequences within each of which the
 transpositions are weakly monotonically increasing  (i.e. $b_i \le b_{i+1}$ for each neighbouring pair).

\et

Evaluating at ${\bf t}_\infty := (1, 0,0,  \dots )$ we have
\be
S_\lambda({\bf t}_\infty) = {1\over h_\lambda}, \quad P_\nu({\bf t}_\infty) = \delta_{\nu, (1)^{|\nu|}}.
\ee
Therefore setting ${\bf s } = {\bf t}_\infty$ in (\ref{hurwitz_hypergeometric_tau_S}) and (\ref{double_power_sum_exp}),
$\tau^{(q, {\bf w}, {\bf z})}({\bf t}, {\bf s})$ restricts to the KP $\tau$-function
\be
\tau^{(q, {\bf w}, {\bf z})} ({\bf t}, {\bf t}_\infty) = \sum_\lambda h_\lambda^{-1} r_\lambda^{(q, {\bf w}, {\bf z})} S_\lambda ({\bf t})
= \sum_{n=0}^\infty q^n  \sum_{\mu, \atop |\mu|=n}  \sum_{{\bf c}\in \Nb^l} 
\sum_{{\bf d}\in \Nb^m}  {\bf w}^{\bf c}   {\bf z}^{\bf d}   F^{\bf c}_{\bf d}(\mu, (1)^n)
P_\mu({\bf t}) ,
\label{KP_reduction}
\ee
where the coefficients $F^{\bf c}_{\bf d}(\mu, (1)^n)$ are the particular values corresponding to no branching at $\infty$. 
We therefore have the following corollary.

\bc{\bf Reduction to KP $\tau$-function. }
\label{single_Hurwitz_KP_reduction}
The KP $\tau$-function $\tau^{(q, {\bf w}, {\bf z})} ({\bf t}, {\bf t}_\infty)$ is the generating function for
the numbers $F^{\bf c}_{\bf d}(\mu, (1)^n)$ counting branched covers satisfying the same conditions
as \autoref{geometric_interpretation}, but with no branch point at $\infty$. Equivalently, they are equal to
the number of paths in the Cayley graph of $S_n$ from an element in the conjugacy class with
cycle type $\mu$ to the identity element (i.e. the number of factorizations of an element in the class $\mu$
as a product of transpositions) that satisfy the conditions of \autoref{combinatorial_interpretation}. 
\ec

 As special cases, consider $(l, m)= (1, 0)$ and $(0,1)$. For any positive integer  $c \in\Nb_+$,  
and pair of partitions $(\mu, \nu)$ with $ |\mu|=|\nu| = n$, let $F^+_c(\mu, \nu)$ and $F^-_c(\mu, \nu)$ 
denote the number of $n$-sheeted branched  covers of  the Riemann sphere $\Pb^1$, up to automorphisms,  
with Euler characteristic  
\be
\chi = 2 - 2g  =  \ell(\mu) + \ell(\nu) -c,
\label{Riemann_Hurwitz_fixed_genus}
\ee
 having either an even ($F^+_c(\mu, \nu)$)
 or an odd  ($F^-_c(\mu, \nu)$) total number of branch points,  including a pair $(0,\infty)$
 with ramification profiles $(\mu, \nu)$.

Let $F^c(\mu, \nu)$ and $F_c(\mu, \nu)$ be the composite Hurwitz 
number $F^{{\bf c}}_{{\bf d}}(\mu, \nu)$ with $(l,m)=(1,0)$,  $c_1 = c$,  and $(l,m)=(0,1)$,
$d_1 =c$), respectively.  According to \autoref{combinatorial_interpretation}, $F^c(\mu, \nu)$ is the number
of strictly monotonically increasing products of $c$ transpositions $(a_1 b_1) \cdots (a_c b_c)$ such that,
if $g\in S_n$ is in the conjugacy class with cycle type $\mu$, the product $(a_1 b_1) \cdots (a_c b_c) g$
is in the conjugacy class $\nu$, while  $F_c(\mu, \nu)$ is the number of  products  
having the same property, but which are weakly monotonically increasing.
The following is an immediate consequence of Theorems \ref{geometric_interpretation}
and \ref{combinatorial_interpretation} for these two cases.
\bc {{\bf The cases $(l, m)=(1, 0)$ and $(0,1)$.}}
\label{lm_01}
\begin{description}
\item[(i) $(l, m)=(1, 0)$:]  \hfill 

In this case, there are  at most three branch points, the ones
at $(0,\infty)$ having ramification profiles $(\mu, \nu)$ and a third one, whose profile $\lambda$
has length
\be
\ell(\lambda) = n-c.
\ee
These are  therefore Belyi curves \cite{Z, AC, KZ}. The combinatorial meaning
of $F^c(\mu, \nu)$ is that it equals the number of paths in the Cayley graph of $S_n$ 
consisting of sequences of $c$ strictly monotonically increasing transpositions,
starting at an element in the conjugacy class of cycle type $\mu$ and ending in the class of type $\nu$.  
\item[(ii) $(l, m)=(0, 1)$:]
\be
F_c(\mu, \nu) = (-1)^{n +c} (F^+_c(\mu, \nu) - F^-_c(\mu, \nu)). 
\label{0_1_case}
\ee
Thus, the number of weakly monotonically increasing paths  of transpositions
that lead from an element in the conjugacy class of type $\mu$ to the class of type $\nu$
is equal to the difference between the number of branched covers with an even or
an odd number of branch points, branching profiles $(\mu, \nu)$ at $(0, \infty)$ and
Euler characteristic  given by (\ref{Riemann_Hurwitz_fixed_genus}).
\end{description}
\ec
If  $\infty$ is not a branch point; i.e.,  its profile is  $\nu=(1)^n$,  corresponding 
to the identity class in $S_n $, Corollaries  \ref{single_Hurwitz_KP_reduction} and  
 \ref{lm_01} imply that the number of factorizations of any element  $g \in S_n$ in the  class $\mu$ 
as a product of $c$ strictly  monotonically increasing transpositions is equal to the number of Belyi curves 
with no branching at $\infty$ and $c$ pre-images  of the additional branch point.
The number of factorizations into weakly increasing subsequences of transpositions is
given by the difference (\ref{0_1_case}) between the number of branched covers having an
even or an odd number of branch points, for the Euler characteristic
given by eq.~(\ref{0_1_case}), and  $\nu = (1)^n$.

\br{\bf Further particular cases.}  The coefficients $F^{\bf c}_{\bf d}(\mu, \nu)$ with  $m=0$
 (i.e.~no  branch points  of  type (iii))  and all $c_a = 1$ count those 
coverings that have simple ramification at all the points $\{q_a\}_{a=1, \dots, l}$ in 
group (ii), and hence these coincide with Okounkov's double Hurwitz number \cite{Ok}. In the combinatorial
interpretation, all the subsequences contain a single transposition, therefore there is is no
monotonicity  imposed. Similarly, if  $c_a=0$ for all $a$'s, and  $d_b=1$  for all
$b$'s, (or equivalently, if $l$=0), it follows that there can only be a single  branch point in each colour group, and that
it is simply ramified. This must therefore also equal Okounkov's double Hurwitz number,
up to an overall sign. In the combinatorial interpretation there is again only one transposition in each sequence,
so there is no condition of monotonicity. In fact, we may choose any subset
of the $c_a$'s and all the $d_b$'s to equal $1$, and the other $c_a$'s  to vanish, and the same result holds.

When $l=0$ and $m=1$ we have, combinatorially, a single sequence of 
$d_1$ weakly monotonically increasing transpositions. This was the case considered in \cite{GGN1, GGN2}, 
and the $\tau$-function shown to be identifiable with the Harish-Chandra-Itzykson-Zuber (HCIZ)
matrix integral \cite{IZ} when the expansion parameter is identified as $z = -1/N$
and the flow parameters equated to the trace invariants of a pair of Hermitian matrices.
 The case  $l=1, m=1$ was explained combinatorially in  \cite{GH}, and
  given a matrix model representation. The case when $l$ is arbitrary and $m=0$ was  considered in \cite{AMMN1, AMMN2}, 
and its combinatorial interpretation was given in \cite{GH}, but no general enumerative
 geometric interpretation seems previously to have been provided, except for the case of Belyi curves
 where $l=2$, $m=0$ and $\mu$ is the trivial partition,  which was studied in detail in \cite{Z, AC, KZ}

A further special case can be obtained by choosing $l$ arbitrary, $m=0$, 
and summing over all coverings with  $l$ additional branch points and fixed genus. Letting
\be
c:= \sum_{a=1}^l c_a
\ee
be the sum of the compliments of their ramification lengths  (i.e. the number of pre-images),
the Riemann-Hurwitz formula  (\ref{Riemann_Hurwitz_fixed_genus}) holds.
Let
\be
F^{(c,l)}(\mu, \nu) = \sum_{c_1, \dots, c_l \atop \sum_{a=1}^lc_a =c} F^{(c_1, \dots, c_l)} (\mu, \nu)
\ee
be the total number of branched covers with up to $2+l$ branch points, ramification profiles 
$(\mu, \nu)$ at $(0, \infty)$ and genus given by (\ref{Riemann_Hurwitz_fixed_genus}). 
The specialization of the $\tau$-function $\tau^{(q, {\bf w}, {\bf z})}({\bf t}, {\bf s})$ to the
case $m=0$, $w_1  = \cdots = w_l = w$ gives
\bea
\tau^{(q, (w)^{\otimes l})}({\bf t}, {\bf s}) &\&
= \sum_{\lambda} q^{|\lambda|} (r_\lambda^{(1,w)})^lS_\lambda({\bf t}) S_\lambda({\bf s})  \cr
&\&= \sum_{\lambda } q^{|\lambda|}\sum_{\mu,\nu \atop |\mu = |\nu| =|\lambda|} 
F^{(c,l)}(\mu, \nu) w^c P_\mu({\bf t}) P_\nu ({\bf s}).
\eea
The combinatorial interpretation of $F^{(c,l)}(\mu, \nu)$ is that it equals the number of
$c$-step paths in the  Cayley graph of $S_n$ consisting of $l$ sequential segments
that are strictly monotonically increasing, with segment lengths anywhere between
$1$ and $n$. 

\er

The next two sections provide the  proofs of Theorems \ref{geometric_interpretation}
 and \ref{combinatorial_interpretation}.

\section[ $\tau^{(q, {\bf w}, {\bf z})}{(\bf t, \bf s})$   generating function  for  generalized Hurwitz numbers]
{Geometric interpretation:  generating function  for  generalized, signed Hurwitz numbers}

\subsection{Double power sum expansion of  $\tau^{(q, {\bf w}, {\bf z})}({\bf t}, {\bf s})$}

The Hurwitz number $H_{g_0}(\mu^{(1)}, \dots \mu^{(j)})$, where $|\mu^{(a)}|=n$, $a= 1, \dots , j$ 
is the number of $n$ sheeted branched covers, up to isomorphism,  of a Riemann surface 
 of genus $g_0$,  with $j$ branch points $\{q_1, \dots , q_j\}$, whose ramification profiles are
given by the partitions $\{\mu^{(a)}\}_{a=1, \dots, j}$. 
 The genus of the covering curve is given  by the Riemann-Hurwitz formula
 \be
 g = {1\over 2} \left(\sum_{a=1}^j \ell(\mu^{(a)}) -nj\right) + n( g_0 - 1)+1
 \ee
Hurwitz numbers can be expressed  as sums  over products of the irreducible
 characters  $\chi_\lambda$ of $S_n$ evaluated at the  conjugacy classes corresponding
  to the partitions using Frobenius' formula, \cite{Frob},  \cite[Appendix~A]{LZ} (see also \cite{L}):
\be
H_{g_0}(\mu^{(1)}, \dots \mu^{(j)}) = \sum_{\lambda, \ |\lambda|=n}
 h_\lambda^{j +2g_0 -2} \prod_{a=1}^j{ \chi_\lambda(\mu^{(a)}) \over Z_{\mu^{(a)}}}.
 \label{frobenius_hurwitz_general}
\ee
We only consider the case where the base curve is the Riemann sphere, so
$g_0 =0$ and  (\ref{frobenius_hurwitz_general}) becomes

\be
H(\mu^{(1)}, \dots \mu^{(j)} )= \sum_{\lambda, \ |\lambda|=n}
 h_\lambda^{j -2} \prod_{a=1}^j{ \chi_\lambda(\mu^{(a)}) \over Z_{\mu^{(a)}}}. 
  \label{frobenius_hurwitz}
  \ee
  
  In order to apply this, we first express the content product formula (\ref{content_product_q_c_d}) for 
  $r_\lambda^{(q, {\bf w}, {\bf z})}$   in terms of the extended  Pochhammer symbol for partitions:
  \be
  (u)_\lambda := \prod_{i=1}^{\ell(\lambda)} (u -i +1)_{\lambda_i}, \quad (u)_i = \prod_{i=0}^{i-1} (u+i). 
  \ee
  Let 
  \be
  u_a := {1\over w_a}, \quad v_b := - {1\over z_b}.
  \ee
  Then
  \bea
  \prod_{a=1}^l \prod_{(i\, j) \in \lambda}(1+w_a(j-i))&\& = \prod_{a=1}^l {(u_a)_\lambda \over u_a^{|\lambda|}} \cr
    \prod_{b=1}^m \prod_{(i\, j) \in \lambda}(1-z_b(j-i))&\& = \prod_{b=1}^m {(v_b)_\lambda \over(v_a)^{|\lambda|}}. 
  \eea
  The Pochhammer symbols may be written in terms special evaluations 
  of Schur functions \cite{Mac, OrSc} as
  \be
  (u)_\lambda = {S_\lambda({\bf t}(u)) \over  S_\lambda({\bf t}_\infty) } = h_\lambda S_\lambda({\bf t}(u))
  \ee
  where
  \be 
  {\bf t}(u) := (u, u/2, \dots, u/i, \dots), \quad  {\bf t}_\infty := (1, 0,0, \dots ),
  \ee
  while the power sum symmetric functions  at these values  are given by
  \be
  P_\mu({\bf t}(u)) =  u^{\ell(\mu)},  \quad P_\mu({\bf t}_\infty) = \delta_{(\mu, 1^{|\mu|})}
  \ee
  Using the Frobenius character formula (\ref{frobenius_character}), this  gives
  \be
  (u)_\lambda = h_\lambda \sum_{\mu, |\mu| = |\lambda|} {\chi_\lambda(\mu) \over Z_\mu} u^{\ell(\mu)}, 
  \ee
  and hence
  \bea
  \prod_{a=1}^l \prod_{(i\, j) \in \lambda}(1+w_a(j-i)) &\& = \prod_{a=1}^l \left(h_\lambda \sum_{\mu, |\mu| = |\lambda|} {\chi_\lambda(\mu) \over Z_\mu} w_a^{\ell^*(\mu)} \right),\\
  {} &\& \cr
   \prod_{b=1}^m \prod_{(i\, j) \in \lambda}(1-z_b(j-i))&\& =\prod_{b=1}^m \left(h_\lambda \sum_{\nu, |\nu| = |\lambda|} {\chi_\lambda(\nu) \over Z_\nu}(- z_b)^{\ell^*(\nu)}\right), 
\eea
where 
\be
\ell^*(\mu) := |\mu| - \ell(\mu),  \quad \ell^*(\nu) := |\nu| - \ell(\nu)
\ee
are the complements of the lengths. Substituting  into the content product formula gives
\be
r_\lambda^{(q, {\bf w}, {\bf z})} = q^{|\lambda|}\prod_{(i,j)\in \lambda }
\left( {\prod_{a=1}^l (1+w_a (j-i)) \over \prod_{b=1}^m (1-z_b(j-i))}\right)
= {\prod_{a=1}^lh_\lambda \sum_{\mu, |\mu| = |\lambda|} {\chi_\lambda(\mu) \over Z_\mu} w_a^{\ell^*(\mu)}  \over \prod_{b=1}^m h_\lambda \sum_{\nu, |\nu| = |\lambda|} {\chi_\lambda(\nu) \over Z_\nu}(- z_b)^{\ell^*(\nu)}}. 
\label{multimonotonic_frobenius_content_product}
\ee
In order to expand as a power series in the parameters ${\bf w}= (w_1, \dots w_l)$, $ {\bf z}= (z_1, \dots z_m)$,
we express the factors appearing in the denominator as
\be
h_\lambda \sum_{\nu, |\nu| = |\lambda|} {\chi_\lambda(\nu) \over Z_\nu}(- z_b)^{\ell^*(\nu)}
= 1 + h_\lambda {\large{ \textstyle \sum'}}_{\nu, |\nu| = |\lambda|}{\chi_\lambda(\nu) \over Z_\nu}(- z_b)^{\ell^*(\nu)}
\ee
where $ \sum'_{\nu, |\nu| = |\lambda|} $ means the sum with the identity class $\nu = 1^{|\lambda|}$ omitted.
Expanding each denominator  factor  gives
\be
{1\over  1+ h_\lambda \large{ \textstyle \sum'}_{\nu, |\nu| = |\lambda|}{\chi_\lambda(\nu) \over Z_\nu}(- z_b)^{\ell^*(\nu)}}
= \sum _{j_b=0}^\infty(-1)^{j_b}
 \left(h_\lambda  \large{ \textstyle\sum'_{\nu, |\nu| = |\lambda|}} {\chi_\lambda(\nu) \over Z_\nu}(- z_b)^{\ell^*(\nu)}\right)^{j_b}.
 \label{denominator_factors}
\ee

\subsection{Signed counting of constrained  branched covers}

We now combine  the computations of the previous subsection for the content
product formula expression $r_\lambda^{(q, {\bf w}, {\bf z})}$  with the Frobenius 
character formula (\ref{frobenius_character}) to expand the $\tau$-function (\ref{hurwitz_hypergeometric_tau_P})
in a series consisting of products of pairs of power sum symmetric functions with coefficients that are Taylor series in the
parameters $(q, w_1, \dots, w_l, z_1, \dots, z_m)$.
Substituting  the content product formulae (\ref{multimonotonic_frobenius_content_product}) 
together with the denominator expansions (\ref{denominator_factors}) into (\ref{hurwitz_hypergeometric_tau_P}),
applying the Frobenius formula (\ref{frobenius_hurwitz}) for the Hurwitz numbers, and combining all 
monomial terms in like powers of $w_a$'s and $z_b$'s,  we obtain
\be
\tau^{(q, {\bf w}, {\bf z})} ({\bf t}, {\bf s}) =
\sum_{n=1}^\infty q^n \sum_{\mu, \nu, |\mu| =|\nu|=n} \sum_{{\bf c} \in \Nb^l } \sum_{{\bf d}\in (\Nb^+)^m}
F^{\bf c}_{\bf d}(\mu, \nu)  {\bf w}^{\bf c} {\bf z}^{\bf d }P_\mu({\bf t} ) P_\nu ({\bf s}),
\label{tau_qwz_hurwitz}
\ee
where multi-index notation has been used:
\be
 {\bf w}^{\bf c} := \prod_{a=1}^l w_a^{c_a}, \quad   {\bf z}^{\bf d} := \prod_{b=1}^m z_b^{c_b}
\ee
and
\bea
F^{\bf c}_{\bf d}(\mu, \nu)\ &\&=  (-1)^{mn + D }  \quad \mathclap{\sum_{\substack{\mu^{(a)}, \ \ell^*(\mu^{(a)})=c_a 
\\ \nu^{(b, i_b)},\  \sum_{i_b=1}^{j_b}\ell^*(\nu^{(b, i_b)}=d_b}}}
\quad\quad  \quad  \sum_{j_1=1}^{d_1} \cdots \sum_{j_m=1}^{d_m} 
\sum_{i_1=1}^{j_1}\cdots \sum_{i_m=1}^{j_m}
(-1)^C H(\mu, \nu,\{ \mu^{(a)}\}, \{{\nu^{(b,i_b)}\}) }, \cr
&\&
\label{signed_coloured_multihurwitz}
\eea 
where
\be
 C= \sum_{b=1}^m j_b, \quad  
 \ee
is the total number of coloured branch points and
\be
D = \sum_{b=1}^m d_b =\sum_{b=1}^m \sum_{i_b=1}^{j_b} \ell^{*}(\nu^{(b, i_b)})
\ee
is the sum of the complements of their ramification profile weights,  which proves \autoref{geometric_interpretation}.

\br Note that  because of the constraints
\be
 \ell^*(\mu^{(a})=c_a , \quad 
\sum_{i_b=1}^{j_b}\ell^*(\nu^{(b, i_b)})=d_b 
\ee
and the fact that all partitions  have the fixed  weight
\be
|\mu^{(a)}| = |\nu^{(b, i_b)}| =|\mu| = |\nu| =n, 
\ee
the number of terms in the sum (\ref{tau_qwz_hurwitz}) is finite.
\er

\section[Combinatorial interpretation: paths in the Cayley graph  ]
{Combinatorial interpretation:  multimonotonic paths in the Cayley graph of $S_n$}

\subsection{The $\{C_\mu\}$ and $\{F_\lambda\}$ bases for the center $\Zb(\Cb[S_n])$ }

The following is a brief version of the method developed in ref.~\cite{GH}, to which the reader is referred 
for further details.
 We recall two standard bases  for the center $\Zb(\Cb[S_n])$ of the group algebra, both labelled
 by partitions of weight $n$. The first  consists of the cycle sums, defined by
\be
C_\mu = \sum_{g\in cyc_\mu}g, \quad   \text{ for } |\mu|=n ,
\ee
where $ cyc_\mu$ denotes the conjugacy class with cycle type $\mu$.
The second consists of the orthogonal idempotents, which may be defined as
\be
F_\lambda := h_\lambda ^{-1} \sum_{\mu, |\mu|=|\lambda|} \chi_\lambda(\mu) C_\mu.
\label{F_lambda_def}
\ee
These satisfy the relations
\be
F_\lambda F_\mu = \delta_{\lambda \mu} F_\lambda.
\ee
from which it follows that these are eigenvectors under multiplication by any element of the
center  $\Zb(\Cb[S_n])$.
The {\em Jucys-Murphy elements}  \cite{Ju, Mu, DG},  defined as consecutive
sums of transpositions 
\be
\JJ_b := \sum_{a=1}^{b-1} (ab), \quad b=1, \dots, n
\label{jucys_murphy}
\ee
 generate a commutative subalgebra of  $\Cb[S_n]$. Moreover, if $G \in \Lambda$ is
a symmetric function, the substitution of the $\JJ_a$'s for the indeterminants
gives an element of the center $\Zb(\Cb[S_n])$
\be
G(\JJ) := G(\JJ_1, \dots , \JJ_n)  \in \Zb(\Cb[S_n])
\ee
whose eigenvalues, under multiplication of $F_\lambda$,  are equal to
the evaluation of $G$ on the content of the Young diagram of the partition $\lambda$
\be
G(\JJ)F_\lambda = G(\cont(\lambda)) F_\lambda,
\ee
where $\cont(\lambda)$ denotes the $n$ element set consisting of the numbers
$j-i$, where $(i,j) \in \lambda$ are the locations of the boxes in the Young diagram. 

\subsection{Multimonotonic paths in the Cayley Graph}

In particular, if we choose $G$ to be the generating functions of the elementary and
complete symmetric functions
\bea
E(w, \JJ) &\&= \prod_{a=1}^n (1+ w \JJ_a) = \sum_{j=0}^n e_j(\JJ) w^j 
\label{generating_functionE}\\
H(z, \JJ) &\&= \prod_{a=1}^n (1-z \JJ_a)^{-1} = \sum_{j=0}^\infty h_j(\JJ) w^j, 
\label{generating_functionH}
\eea
we obtain
\bea
E(w, \JJ) F_\lambda &\&= \prod_{(ij)\in \lambda} (1 + w (j-i)) F_\lambda \\
H(z, \JJ) F_\lambda &\&= \prod_{(ij)\in \lambda} (1 - z (j-i))^{-1} F_\lambda.
\eea
Combining these by multiplication gives
\be
q^{|\lambda|} \prod_{a=1}^l \prod_{b=1}^m E(w_a, \JJ) H(z_b, \JJ) F_\lambda
= r_\lambda^{(q, {\bf w}, {\bf z})} F_\lambda.
\ee
On the other hand,  the expansions (\ref{generating_functionE}), (\ref{generating_functionH}),
together with  the fact that
\bea
e_i(\JJ) &\&= \sum_{(j_1, \dots, j_i) \ss (1, \dots, n) \atop j_1 < j_2 \dots < j_i} \JJ_{j_1} \cdots \JJ_{j_i}
\label{e_i}\\
f_i(\JJ)&\&  = \sum_{(j_1, \dots, j_i) \ss (1, \dots, n) \atop j_1 \le j_2 \dots \le j_i} \JJ_{j_1} \cdots \JJ_{j_i}
\label{f_i}
\eea
and the definition (\ref{jucys_murphy}) of the Jucys-Murphy elements implies that if this
same element is  applied to the cycle sums, we obtain
\be
q^{|\lambda|}\prod_{a=1}^l \prod_{b=1}^m E(w_a, \JJ) H(z_b, \JJ)  C_\mu
= q^{|\lambda|} \sum_{\nu, |\nu|=|\mu|} \tilde{F}^{\bf c}_ {\bf d}(\mu, \nu)  {\bf w}^{\bf c} {\bf z}^{\bf d}C_\nu,
\label{cycle_sum_action}
\ee
where $ \tilde{F}^{\bf c}_ {\bf d}(\mu, \nu)$ is the number of products of  $k$ transpositions $(a_1 b_1) \cdots (a_{k} b_{k})$,
\be
k = \sum_{a=1}^l c_a + \sum_{b=1}^m  d_b
\ee
satisfying
\be
C_\nu = (a_1 b_1) \cdots (a_k b_k) C_\mu
\ee
 such that these may be grouped into successive sequences, corresponding to each of the factors in the product
$\prod_{a=1}^l \prod_{b=1}^m E(w_a, \JJ) H(z_b, \JJ)$, expanded in powers
of $w_a$ and $z_b$. These consist first of a sequence of $l$ bands of transpositions  having lengths $c_a$, $ a=1, \dots, l$
 that are strictly monotonically increasing in the second factors of $(a_i \ b_i)$,   $b_i < b_{i+1}$, followed by $m$ bands 
 of lengths $d_b$, $b=1, \dots ,m$,  in which they are weakly monotonically increasing.
 
 Substituting the change of basis formula (\ref{F_lambda_def}) into (\ref{cycle_sum_action}) and equating
 the coefficients in the sums over the $F_\lambda$ basis gives:
 \be
 \chi_\lambda(\mu)  r_\lambda^{(q,{\bf w}, {\bf z})} =  q^{|\lambda|} \sum_{\nu,  |\nu|=|\mu|
 = |\lambda|} Z_\nu\chi_\lambda(\nu)\sum_{{\bf c}\in \Nb^l} 
\sum_{{\bf d}\in \Nb^m}  \tilde{F}^{\bf c}_ {\bf d}(\mu, \nu) {\bf w}^{\bf c} {\bf z}^{\bf d}. 
 \ee
 By the orthogonality of group characters
 \be
 \sum_{\lambda ,  |\lambda|=|\mu|=|\nu|} \chi_\lambda(\mu) \chi_{\lambda}(\nu)= Z_\mu \delta_{\mu \nu}
 \ee
 this is equivalent to
 \be 
 \sum_{\lambda, |\lambda|=n}  r_\lambda^{(q, {\bf w}, {\bf z}) } \chi_\lambda (\mu) \chi_\lambda(\nu)
 = q^n Z_\mu Z_\nu \sum_{{\bf d}\in \Nb^m} \tilde{F}^{\bf c}_ {\bf d}(\mu, \nu) {\bf w}^{\bf c} {\bf z}^{\bf d}.
 \ee
 Therefore, by eqs.~(\ref{hurwitz_hypergeometric_tau_P}),  (\ref{double_power_sum_exp}) we have 
 \be
 \tilde{F}^{\bf c}_ {\bf d}(\mu, \nu) = F ^{\bf c}_ {\bf d}(\mu, \nu) ,
 \ee
 which completes the proof of \autoref{combinatorial_interpretation}. 
 
\br
It follows from the results of \cite{KMMM} and \cite[Appendix~A]{HOr1} that all these generating
functions have representations as matrix integrals. They therefore also satisfy Virasoro constraints,
and  their multitrace resolvent correlators may be computed  through the methods
of topological recursion \cite{EO}.
\er

\section[Fermionic representation of $\tau^{(q, {\bf w}, {\bf z})}{(\bf t, \bf s})$] 
{Fermionic representation of $\tau^{(q, {\bf w}, {\bf z})}{(\bf t, \bf s})$}
\label{fermionic_representation}

Finally, we give the fermionic representation of the 2D Toda $\tau$-functions
$\tau^{(q, {\bf w}, {\bf z})}{(\bf t, \bf s})$, following \cite{ DJKM, Takeb, OrSc, HOr1, HOr2}.
Any chain of 2D Toda $\tau$-functions  of hypergeometric type may be represented, 
in the fermionic Fock space approach, as vacuum state expectation values of the form
\be
\tau_r (N, {\bf t}, {\bf s}) = \langle N |  \hat{\gamma}_+({\bf t}) \hat{C}_\rho \hat{\gamma}_-({\bf s}) |  N \rangle, 
\ee
where  $\langle N|$, $|N \rangle$ denote the left and right vacuum vectors in the $N$th charge sector,
\be
\hat{\gamma}_+({\bf t}) := e^{\sum_{i=1}^\infty t_i J_i}, \quad \hat{\gamma}_-({\bf s})
 ;= e^{\sum_{i=1}^\infty s_i J_{-i}},   \quad  \text{where }J_i := \sum_{j\in \Zb}  \psi_j \psi^\dag_{i+ j}, \quad i \in \Zb,
\ee
are the generators of the two infinite abelian Toda flows,   $\{\psi_i$, $\psi_i^\dag$, $i \in \Zb\}$ are Fermi creation and 
annihilation operators that  satisfy the usual anticommutation relations and vacuum annihilation conditions
\be
[\psi_i, \ \psi^\dag_j]_+ = \delta_{ij}, \quad \psi_i | N\rangle =0 \  \text{ if } \  i<N, 
\quad \psi_i^\dag | N \rangle = 0 \   \text{ if } \ i \ge N
 \ee
 and
 \be
 \hat{C}_\rho = e^{\sum_{j\in \Zb} T_j :\psi_j \psi_j^\dag:} 
 \ee
is  an element of the infinite abelian group of diagonal elements that generate
 {\em generalized  convolution flows}  \cite{HOr2}. 
  The orthonormal Fermionic Fock basis states $|\lambda; N\rangle$ are labelled by pairs $(\lambda, N)$ 
consisting of a partition  $\lambda$ and an integer $N$.
 \be
 |\lambda; N \rangle := (-1)^{\sum_{i=1}^r b_i}\prod_{i=1}^r \psi_{a_i +N} \psi^\dag_{-b_i-1 +N} |N\rangle
 \ee
 where the partition $\lambda$, expressed in Frobenius notation \cite{Mac},  is $(a_1, \dots, a_r | b_1, \dots , b_r)$.

 The double Schur function expansion (\ref{hypergeometric_tau_expansion})   follows from the 
 fact that  Schur  functions have the following fermionic matrix element expressions
 \be
S_\lambda({\bf t})= \langle \lambda; N | \hat{\gamma}_-({\bf t})| N \rangle  = \langle N | \hat{\gamma}_+({\bf t})| \lambda; N \rangle
 \ee
  which, in turn, follow from Wick's theorem.
 Defining
 \be 
\rho_j :=  e^{T_j}, 
 \ee
 it follows that  the basis vectors $|\lambda; N\rangle$ are eigenvectors of the convolution flow group elements
 \be
  \hat{C}_\rho |\lambda; N\rangle = r_\lambda(N) |\lambda;  N \rangle,
  \ee
with eigenvalues $r_\lambda(N)$  given by the content product formula (\ref{r_content_product}).
Inserting a sum over a complete set of intermediate states gives the double Schur function expansion  (\ref{hypergeometric_tau_expansion}) for
 $\tau_r(N, {\bf t}, {\bf s})$. 
  
  The particular case  (\ref{hurwitz_hypergeometric_tau_S}) corresponding to our
  family   $\tau^{(q, {\bf w}, {\bf z}) }(N, {\bf t}, {\bf s})$ of 
  generating functions  is obtained by choosing the parameters $\rho_j$ 
   to be $\rho_j^{(q, {\bf w}, {\bf z})}$,  as defined in (\ref{rho_q_w_z_def}). The corresponding
   values of the convolution flow parameters are 
   \bea
T_j^{(q, {\bf  w}, {\bf z}) } &\& := j \ln q +\sum_{k=1}^j   \sum_{a=1}^l \ln(1+ k w_a)- \sum_{k=1}^j \sum_{b=1}^m \ln \left(  1- k z_b\right)
\quad  \text{for}\  j>0, \quad  \cr
 T^{(q, {\bf  w}, {\bf z})}_0 &\& =0, 
  \cr
T_{-j}^{(q, {\bf  w}, {\bf z}) } &\& := - j \ln q +  \sum_{k=0}^{j-1} \sum_{b=1}^m \ln(1+ k z_b)   -\sum_{k=0}^{j-1}   \sum_{a=1}^l \ln ( 1- k w_a) \quad  \text{for}\  j>0, 
\eea
which gives
   \be
  \tau_r(N, {\bf t}, {\bf s})  =  \tau^{(q, {\bf w}, {\bf z})}(N, {\bf t}, {\bf s}).
   \ee
   \br
Besides the $N=0$ sector of the fermionic Fock space, with orthonormal
basis $\{|\lambda;0\rangle\}$,  on which the group of generalized convolution flows acts diagonally,
and  the  direct sum  $\oplus_{n=0}^\infty \Zb (\Cb[S_n])$,  where a corresponding infinite group 
 acts diagonally, with the same eigenvalues,  on the basis $\{F_\lambda\}$, as detailed in ref.~(\cite{GH}),
there is also the bosonic Fock space representation, in which the $\tau$-function is viewed
as a symmetric function of an infinite number of bosonic variables, with orthonormal basis 
given by the Schur functions $\{S_\lambda\}$. A corresponding infinite abelian group of 
operators acts diagonally in this space, with eigenvalues also given by a content product formula. 
Their infinitesimal generators are expressible as differential operators in the flow parameters  $(t_1, t_2, \dots)$
 with polynomial coefficients, referred to sometimes as {\em cut and join operators} \cite{GJ, AMMN2, MMN}. 
 These may be viewed as generating an abelian group  through exponentiation (i.e. the solution of
a diffusion-like equation),  with  a vacuum $\tau$-function as initial condition.
But these are not, in general,  symmetries of the 2D Toda hierarchy and the solutions are
 not  necessarily 2D Toda $\tau$-functions, even though they admit series representations as 
 diagonal sums over products of Schur functions.  Only the smaller group,  consisting of 
operators acting as generalized convolution flows \cite{HOr2}, whose eigenvalues are 
of the  content product form,  give rise to hypergeometric $\tau$-functions.
\er

 \medskip
\noindent {\it Acknowledgements.} The authors would like to thank Leonid Chekhov,  Mathieu Guay-Paquet, 
 Maxim Kazarian, Sergei Lando,   Andrei Mironov and Sergei Natanzon  for helpful  discussions 
 and comments.

\bigskip
 
\bigskip \bigskip

\newcommand{\arxiv}[1]{\href{http://arxiv.org/abs/#1}{arXiv:{#1}}}

\bigskip
\noindent

\end{document}